\documentclass[prl,twocolumn]{revtex4}
\usepackage{color,graphicx,shortvrb}

\begin{document}
\title{Atom--Molecule Coherence in a Bose-Einstein
Condensate}

\author
{Elizabeth~A. Donley,  Neil~R. Claussen, Sarah~T. Thompson, and
Carl~E. Wieman}

\affiliation{JILA, University of Colorado and National Institute
of Standards and Technology, Boulder, Colorado 80309-0440}
\date{\today}

\begin{abstract}
Coherent coupling between atoms and molecules in a Bose-Einstein
condensate (BEC) has been observed. Oscillations between atomic
and molecular states were excited by sudden changes in the
magnetic field near a Feshbach resonance and persisted for many
periods of the oscillation. The oscillation frequency was measured
over a large range of magnetic fields and is in excellent
quantitative agreement with the energy difference between the
colliding atom threshold energy and the energy of the bound
molecular state. This agreement indicates that we have created a
quantum superposition of atoms and diatomic molecules, which are
chemically different species.
\end{abstract}

\maketitle \vspace{1 cm}

There is considerable interest in extending the applications of
ultracold atoms to ultracold molecules. One route for producing a
very cold and possibly Bose-condensed sample of molecules is to
create the molecules from an atomic BEC. Wynar et al.$^{1}$
created cold $^{87}$Rb$_2$ molecules in a single ro-vibrational
state of the ground-state potential from an $^{87}$Rb BEC using a
two-photon stimulated Raman transition. The authors could not
probe the coherence properties of the molecules in that state, but
the prospect of creating a superposition of atomic and molecular
condensates initiated a flood of theoretical work on the
subject$^{2-6}$. Ultracold molecules have also recently been
formed through photoassociation of a sodium BEC$^7$.

Utilizing the natural atom--molecule coupling that arises from a
Feshbach resonance is an alternate route for producing ultracold
molecules from an atomic BEC, and it is the route we have followed
here. A Feshbach resonance is a scattering resonance for which the
total energy of two colliding atoms is equal to the energy of a
bound molecular state, and atom--molecule transitions can occur
during a collision$^{8-12}$. A schematic representation of the
potentials involved is shown in the inset of Fig. 1A. For our
$^{85}$Rb resonance, BEC atoms in the $F=2$, $m_F=-2$ state
collide on the open-channel threshold. $F$ and $m_F$ are the total
spin and spin-projection quantum numbers. The bound state in the
closed channel differs in energy by an amount $\epsilon$ from the
open-channel threshold. The bound molecular wave function can be
described as a sum of amplitudes of different hyperfine components
($F$, $m_F$) having $M_F = m_{F,1} + m_{F,2} = -4$$^{13}$. Because
of their different spin configurations, the atoms and molecules
generally have different magnetic moments and the difference
depends on magnetic field. Thus $\epsilon$ depends on magnetic
field and the degree of atom--molecule coupling is magnetically
tunable. The energy difference between the free atoms and the
bound molecules is plotted in Fig. 1A. This behavior of the
bound-state energy also causes a resonance in the scattering
length, $a$, which is shown in Fig. 1B. The scattering length
characterizes the mean-field interaction energy of a BEC.

\begin{figure}
\begin{center}
\includegraphics[bb=152 133 468 567, clip,scale=0.7]{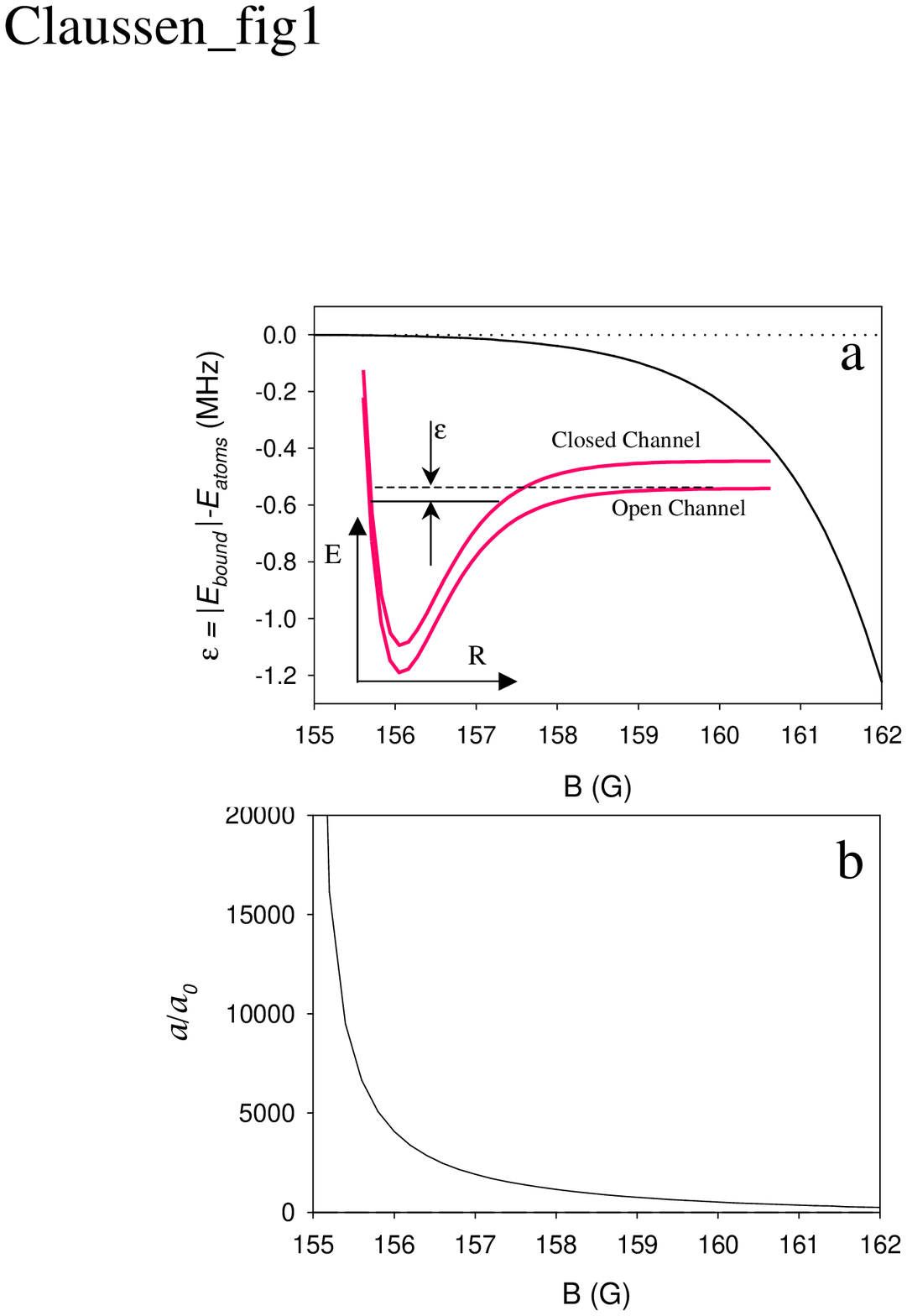}
 \end{center}\caption{
Feshbach resonance bound-state energy and scattering length. (A)
Energy splitting versus magnetic field. The resonance is centered
at $\sim$155~G. The solid curve is a theoretical estimate of the
energy found with a coupled-channels calculation$^{14}$, and the
dotted line indicates $\epsilon = 0$. The inset schematically
shows the collision channels involved in the resonance. $\epsilon$
depends on magnetic field because the atoms and molecules have
different magnetic moments and thus the potentials have different
Zeeman shifts. (B) Scattering length versus magnetic field for
fields above the Feshbach resonance.
 }
 \label{Fig:1}
 \end{figure}

When the magnetic field is tuned to values near the Feshbach
resonance, theory predicts coherent oscillations between the
atomic and molecular states, but there is significant disagreement
on the conversion fraction and the coherence properties$^{16-21}$.

In experiments with a sodium BEC, Stenger et al.$^{22}$ observed
that inelastic losses were dramatically enhanced when they ramped
the magnetic field across the Feshbach resonance. We observed
similar results for $^{85}$Rb, but with lower rates$^{23,24}$. It
is likely that the formation of molecules played a role in the
loss, but there was no experimental evidence for the presence of
molecules and the results followed a loss-rate dependence on time.
More recently, we measured the time dependence of the loss in an
$^{85}$Rb BEC by applying controlled magnetic-field pulses toward
but not across the Feshbach resonance$^{25}$. We observed the
surprising result that under some conditions, shorter, more rapid
pulses actually led to more loss than longer, slower pulses that
spent more time near the resonance. The time dependence of the
loss was suggestive of a nonadiabatic mixing of states, with the
only states within a reasonable energy range being the normal
atomic BEC state and the nearby bound molecular state.

In this work we show that much of the loss is likely due to the
coherent mixing of atomic and molecular states. To create a
superposition and probe its coherence, we applied two short
magnetic-field pulses toward the Feshbach resonance, separated by
a ``free evolution'' time during which the magnetic field was held
at a constant value some distance from the resonance. We measured
the number of atoms in the condensate as a function of time
between the two pulses for various values of the steady-state
magnetic field between the pulses. We observed dramatic
oscillations in the number of atoms remaining in the atomic BEC at
frequencies corresponding to the energy splitting between the
molecular and the atomic states.

{\bf Experimental methods}. The apparatus has been described in
detail elsewhere$^{23,25}$. We first created $^{85}$Rb
condensates$^{23}$ typically containing 16,500 atoms, with fewer
than 1,000 uncondensed thermal atoms. The initial number
$N_{init}$ fluctuated from shot to shot by $\sim$500 atoms ($\sim
3$\% number noise). After producing the condensate at a field of
$\sim$162~G, we ramped the magnetic field adiabatically to
$\sim$166~G, corresponding to an initial scattering length
$a_{init} \simeq 10~a_0$, where $a_0 = 0.053$~nm. The spatial
distribution of the atoms was Gaussian with a peak atom density of
$n_0 = 5.4 \times 10^{13}$~cm$^{-3}$, and the trap frequencies
were (17.4$\times$17.4$\times$6.8)~Hz. After preparing the
condensate we applied a selected fast magnetic-field pulse
sequence by sending an appropriate time-dependent current through
an auxiliary magnetic-field coil$^{25}$. A typical pulse sequence
is shown in Fig. 2. It is composed of two nearly identical short
trapezoidal pulses separated by a region of constant (but
adjustable) magnetic field. Upon completion of the fast-pulse
sequence in Fig. 2, we ramped the magnetic field from $\sim$166~G
to $\sim$157~G in 5~ms and held at that field for an additional
7~ms to allow the repulsive mean-field energy to expand the
condensate. Then we turned off the magnetic trap and used
destructive absorption imaging 12.8~ms later to observe the atomic
condensate and measure the number of remaining atoms$^{26}$. This
detection scheme was neither sensitive to atoms with kinetic
energies larger than $\sim 2~\mu$K nor to atoms in off-resonant
molecular states. We determined the value of the magnetic field
between the pulses, $B_{evolve}$, by measuring the resonance
frequency for transitions from the $F=2$, $m_F = -2$ to the $F=2$,
$m_F = -1$ spin state by applying a 10~$\mu$s RF pulse to a
trapped cloud of atoms$^{25}$.

\begin{figure}
\begin{center}
\includegraphics[bb=212 379 431 532, clip,scale=1.2]{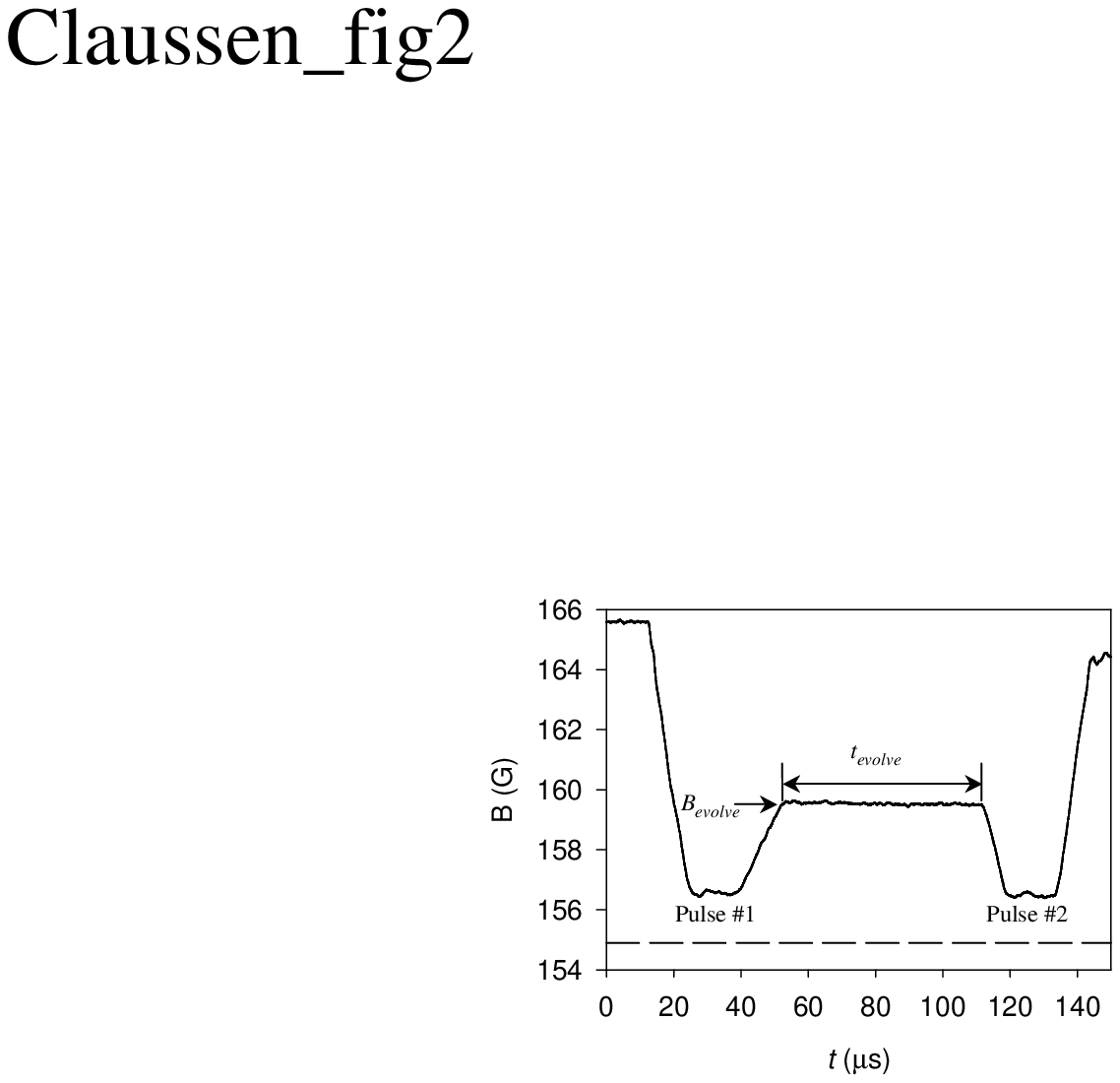}
 \end{center}\caption{
 Magnetic field pulse shape. Fields shown for pulses \#1 and \#2
correspond to scattering lengths of $\sim$2500~$a_0$, and the free
precession field $B_{evolve}$ corresponds to a scattering length
of $\sim$570~$a_0$. The dashed line indicates the position of the
Feshbach resonance. In the text, we refer to the free precession
time as $t_{evolve}$. The rise/fall time for all of the pulses
that we used was 14~$\mu$s.}
 \label{Fig:2}
 \end{figure}

{\bf Burst and remnant oscillations}. As we observed for single
pulses toward the Feshbach resonance$^{25}$, there were two
distinct components of atoms observed in the absorption images and
a third ``missing'' component that we could not detect. One of the
observed components was a cold remnant BEC which was not
noticeably heated or excited by the fast-pulse sequence, while the
other component was a relatively hot ($\sim$150~nK) ``burst'' of
atoms that remained magnetically trapped during the BEC expansion
time. Using a variational approach$^{27}$ to model the mean-field
expansion that we applied to the BEC remnant to measure its
number, $N_{remnant}$, we found that we should impart $\le 3$~nK
worth of energy to the remnant before imaging. This estimate
agrees well with the expansion velocity that we observed after the
trap turn-off. Thus the remnant BEC was nearly $50 \times$ colder
than the burst.

The missing component contained atoms that were in the initial
sample but were not detected after the trap turn-off. To find the
number of atoms in the remnant BEC and the number of burst atoms,
we allowed the magnetic trap to ``focus'' the burst cloud before
imaging$^{28}$. A typical image is shown in Fig. 3. We fit the
focussed burst (which had a much larger spatial extent than the
remnant) with a two-dimensional Gaussian surface, excluding the
central region of the image that contained the remnant. This fit
yielded the number of burst atoms, $N_{burst}$. Subtracting this
fit from the image and performing a pixel-by-pixel sum of the
central region of the image gave $N_{remnant}$.

\begin{figure}
\begin{center}
\includegraphics[bb=150 356 477 511, clip,scale=0.75]{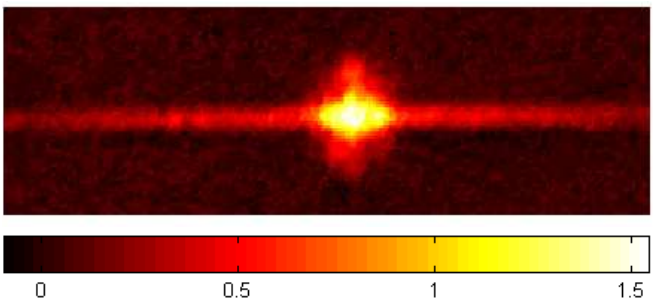}
 \end{center}\caption{
 An absorption image taken after the fast magnetic-field pulse
sequence and the mean-field expansion. The colorbar indicates the
optical density. The horizontal and vertical directions coincide
with the axial and radial axes of the trap, respectively. The
dimensions of the image are $366 \times 52$~$\mu$m. The BEC
remnant is the roughly spherical cloud at the center, while the
burst atoms are focussed into a thin line along the axial
direction. Note the dramatic difference between the two spatial
distributions, owing to the large difference in their mean
energies ($\langle E_{burst} \rangle \simeq 50 \times \langle
E_{remnant} \rangle$).}
 \label{Fig:3}
 \end{figure}

$N_{remnant}$  versus $t_{evolve}$ is plotted in Fig. 4 for two
different values of $B_{evolve}$. The number clearly oscillates.
Changing the value of $B_{evolve}$ affected the oscillation
frequency dramatically (note the change in scale from Fig. 4A to
Fig. 4B). After only pulse \#1 and the subsequent constant field
but with no pulse \#2, $N_{remnant}$ showed no variation except
for a slow decay consistent with the loss rate expected for a
single pulse to that field$^{25}$.

\begin{figure}
\begin{center}
\includegraphics[bb=240 285 431 561, clip,scale=1.2]{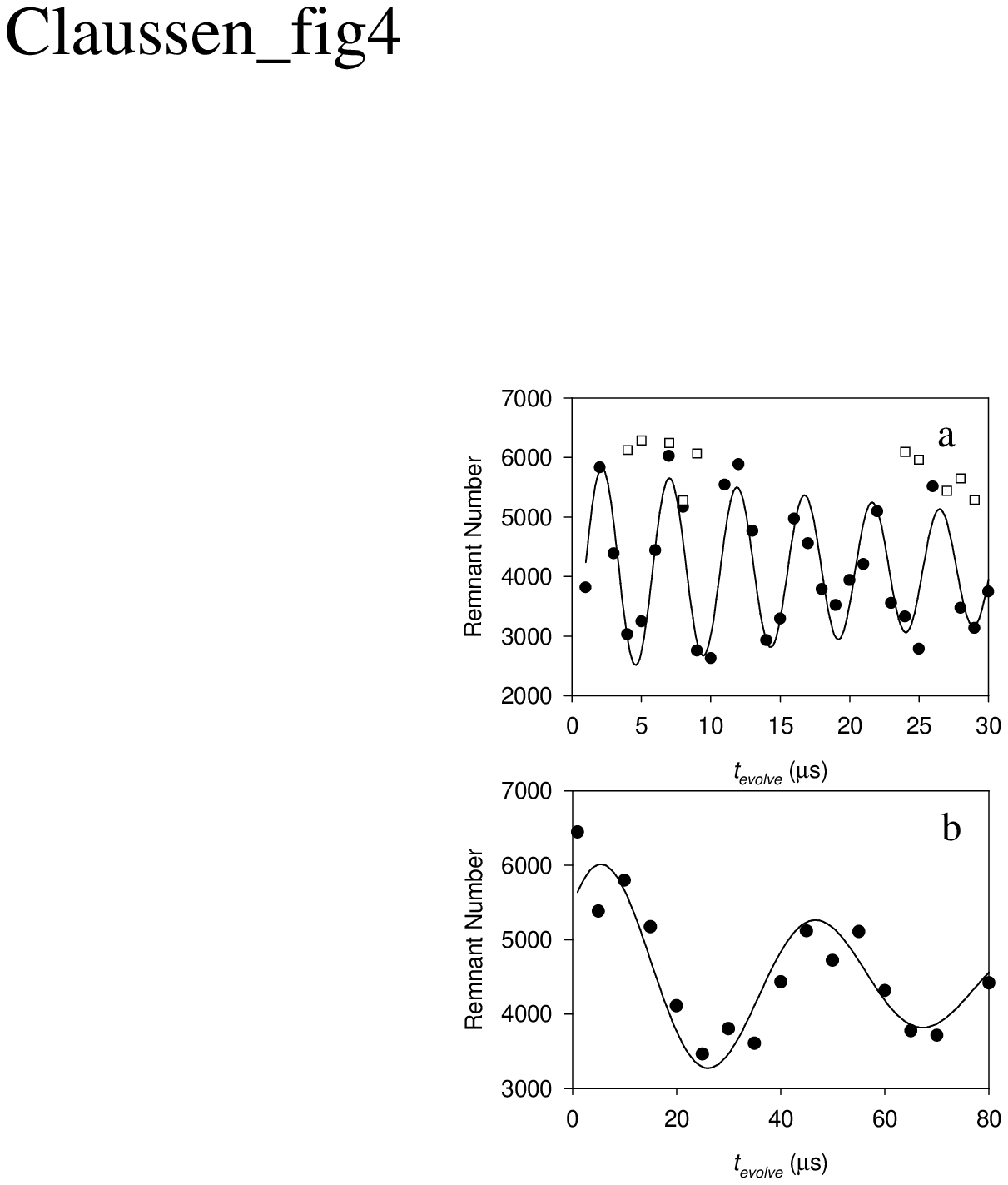}
 \end{center}\caption{
$N_{remnant}$ versus $t_{evolve}$ for $n_0 = 5.4 \times
10^{13}$~cm$^{-3}$. (A) $B_{evolve}$ = 159.69(4)~G ($a_{evolve} =
590~a_0$). The data fit to a damped sine wave with a frequency of
207(2)~kHz and a decay time of 46(21)~$\mu$s. The open squares
near $N_{remnant} = 6000$ indicate the number remaining versus
time after only pulse \#1 and $t_{evolve}$ at 159.69~G. (B)
$B_{evolve}$ = 157.60(4)~G ($a_{evolve} = 1390~ a_0$). Note the
increase in time scale from (A). These data fit to an oscillation
frequency of 23.9(12)~kHz and a decay time of 82(38)~$\mu$s.
 }
 \label{Fig:4}
 \end{figure}

We have taken data similar to those in Fig. 4 for a variety of
different $B_{evolve}$ values.  As in Fig. 4, we fit each curve to
the function $y = y_0 + A \exp{(-t/\tau)} \sin{(2 \pi \nu t +
\phi)}$ to find the oscillation frequency $\nu$ and decay time
constant $\tau$. The measured frequencies are plotted versus
$B_{evolve}$ in Fig. 5 along with theoretical predictions for the
bound-state energy relative to the atomic state.

In the regime where the scattering length is much larger than the
radius of the interatomic potential well, the bound state energy
for an arbitrary attractive potential can be approximated by
$\epsilon = - \hbar^2/m a^2$ $^{29}$. $\hbar$ is Plank's constant
divided by $2 \pi$, $m$ is the atomic mass, and $a$ is the
scattering length. The same equation relates the bound state
energy to the effective scattering length, which is calculated
from the Feshbach resonance parameters through the relation $a =
a_{bg} \times (1 - \frac{\Delta}{B-B_0})$, where $a_{bg}$ is the
background scattering length, $\Delta$ is the width of the
Feshbach resonance, and $B_0$ is the resonant magnetic
field$^{30}$.

The quantity $|\epsilon|/h$ is plotted with no adjustable
parameters in Fig. 5. The measured oscillation frequencies are in
excellent agreement with this simple model over the range of
magnetic fields where the model is expected to be valid. The
theoretical results found with a much more sophisticated
coupled--channels scattering calculation$^{14}$  in Fig. 5 are in
excellent agreement with the data over the entire range.

\begin{figure}
\begin{center}
\includegraphics[bb=162 276 454 491, clip,scale=0.75]{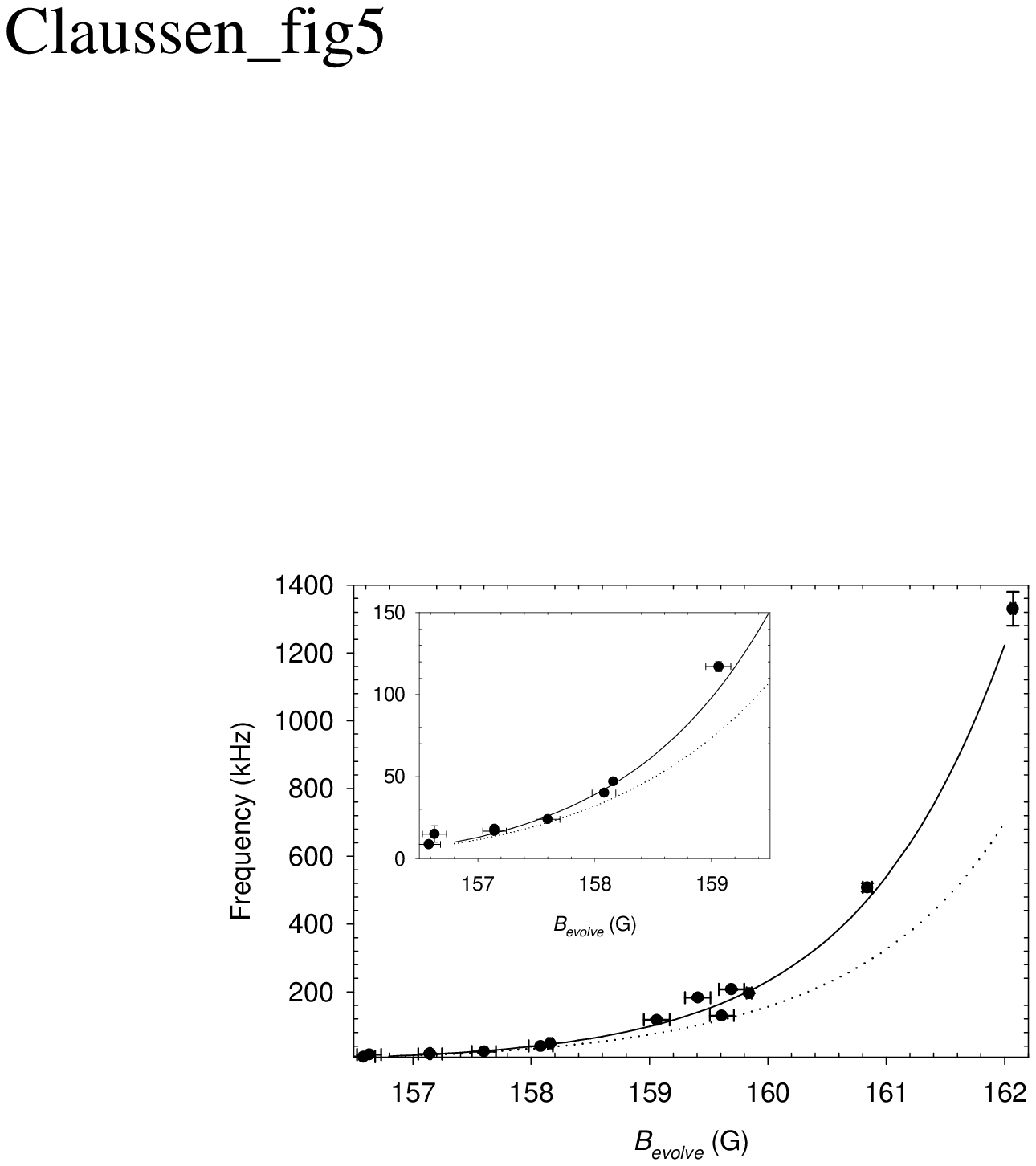}
 \end{center}\caption{
Oscillation frequency versus magnetic field. The points are the
measured frequencies. The solid line is the energy difference
between the atom--atom threshold and the bound molecular state
found by S. Kokkelmans with a coupled-channel scattering
calculation. The dotted line is a plot of $\epsilon/h$. The inset
is an expanded view of the lower-frequency data. The maximum
frequency that we could measure was limited only by timing jitter
and finite resolution in the experiment. The magnetic-field
measurements for the points with the smallest horizontal error
bars were performed on the same days as the corresponding
frequency measurements. The error bars for the points with larger
field uncertainties were inflated by 100~mG to account for
estimated day-to-day field drifts. }
 \label{Fig:5}
 \end{figure}

The fact that the oscillations occurred at exactly the frequency
corresponding to the bound-state energy clearly indicates that we
are creating a coherent superposition of atoms and molecules with
the sudden magnetic-field pulses$^{11,12}$. Although we do not
have a detailed understanding of how the field pulses couple atoms
and molecules, by choosing the shapes of the perturbing pulses
such that a single pulse results in roughly 50\% loss, we observe
high-contrast oscillations in the number of atoms in the atomic
BEC.

From the amplitude of the oscillations, one can put a lower bound
on the number of molecules being created. Take, for example, the
data in Fig. 4A. The amplitude of the atom oscillations was
1800(300) atoms. Assuming the fringes are coming from interference
with molecules, there must be at least 1800/2 = 900(200) molecules
on average. Assuming that the missing atoms are molecules that we
fail to convert back into atoms gives an upper bound of 3200(100)
molecules for the conditions of Fig. 4A.

The damping time for the oscillations, $\tau_{decay}$, was more
difficult to measure with high precision than the oscillation
frequency. To within our measurement precision, $\tau_{decay}$ did
not depend on $B_{evolve}$, but our uncertainties in
$\tau_{decay}$ were as large as 100\% for some fields. We had the
highest precision measurements for frequencies around 200~kHz
where the oscillation period was long compared to our experimental
timing jitter but short compared to $\tau_{decay}$. At frequencies
near 200~kHz, we measured $\tau_{decay} = 38(8)~\mu$s for $n_0 =
1.3 \times 10^{14}$~cm$^{-3}$ and $\tau_{decay} = 91(33)~\mu$s
when we decreased $n_0$ to $1.1 \times 10^{13}$~cm$^{-3}$.

$N_{burst}$ also had interesting dependencies. For the conditions
under which most of the data were collected, the burst contained
$\sim$5000 atoms on average, which is $\sim$30\% of $N_{init}$.
$N_{burst}$ depended on density and varied from one-half of the
atoms lost from the condensate for our typical peak density of
$n_0 = 5.4 \times 10^{13}$~cm$^{-3}$ to nearly all of the atoms
lost from the condensate for $n_0 = 1.1 \times 10^{13}$~cm$^{-3}$.
$N_{burst}$, $N_{remnant}$, and total number of atoms detected are
plotted in Fig. 6 for $B_{evolve}$ = 159.84(2)~G and $n_0 = 1.1
\times 10^{13}$~cm$^{-3}$ ($5 \times$ lower density than was used
for the data shown in Fig. 4). All three components oscillated at
the same frequency. The burst oscillation lagged behind the
remnant oscillation by 155(4)$^\circ$. Since the relative phase
shift is nearly 180$^\circ$, the oscillation amplitude for the
total number was smaller than either the burst or the remnant
oscillation amplitudes. The relative phase depended sensitively on
the fall time of pulse \#2. For example, when we increased the
fall time from 11~$\mu$s to 159~$\mu$s, the burst oscillation then
lagged behind the remnant oscillation by 68(7)$^\circ$ and the
peak-to-peak amplitude of the total number oscillation was
5600(400).

 \begin{figure}
\begin{center}
\includegraphics[bb=227 398 421 535, clip,scale=1.2]{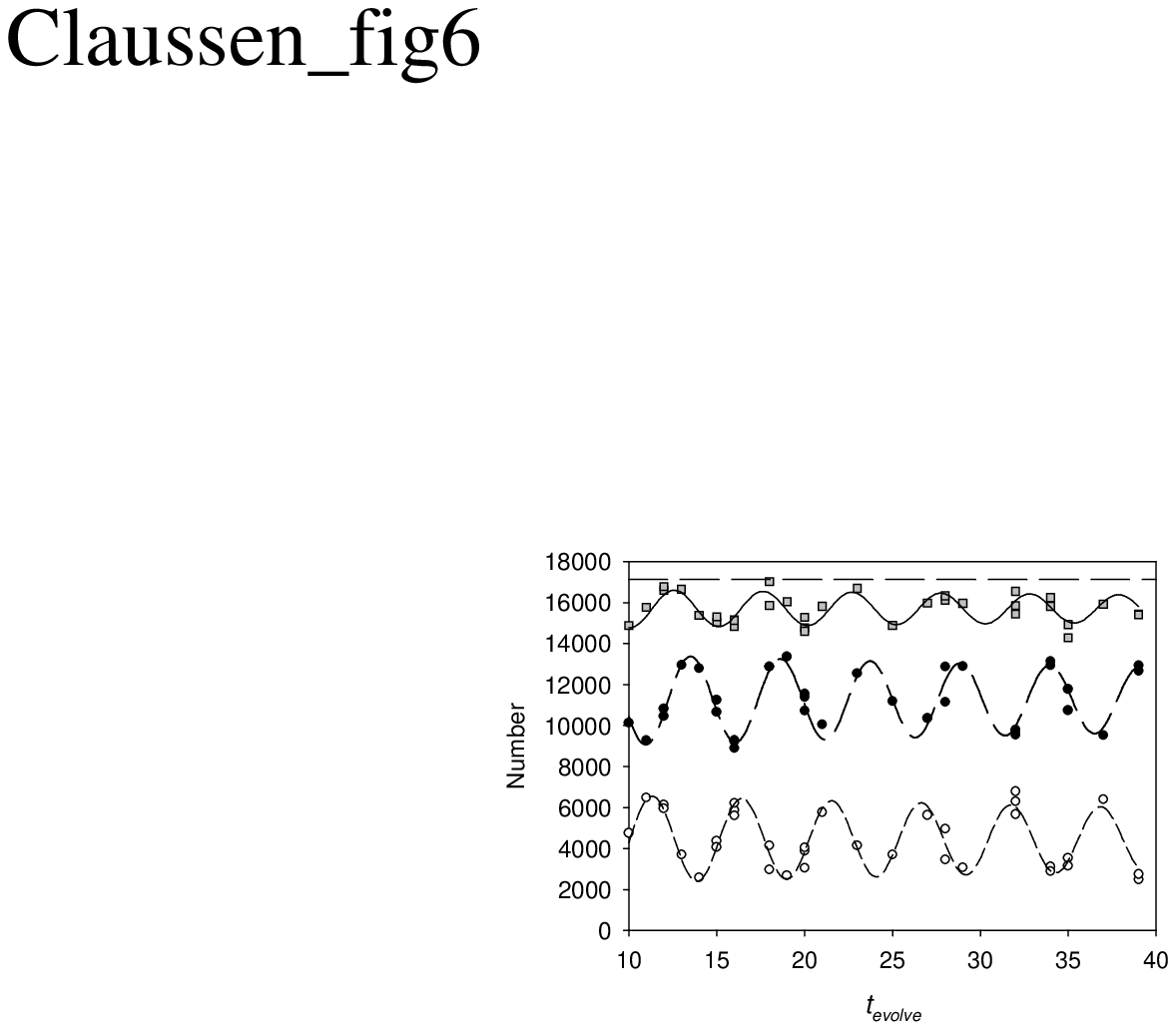}
 \end{center}\caption{
 Number versus $t_{evolve}$ for $n_0 = 1.1 \times
10^{13}$~cm$^{-3}$. From bottom to top, the data are plots of
$N_{burst}$ (open circles), $N_{remnant}$ (filled circles), and
the total number of observed atoms (gray squares). Each data set
was fit to a damped sine wave resulting in the displayed fits.
$N_{init}$ = 17,100 is indicated by the flat dashed line.
$B_{evolve}$ = 159.84(2)~G and the remnant data fit to an
oscillation frequency of 196(1)~kHz and $\tau_{decay}$ =
91(33)~$\mu$s. To produce condensates with the lower density for
these measurements, the fast-pulse sequence was applied from our
evaporation field of 162.2~G and the amplitudes of pulses \#1 and
\#2 were reduced to $\sim$7~G.}
 \label{Fig:6}
 \end{figure}

For the conditions of Fig. 6, $N_{init}$ = 17,100 exceeded the
time-averaged total number of atoms counted after the pulse
sequence by 8(3)\% on average. For the higher-density measurements
in Fig. 4, 39(4)\% of the atoms were missing. Experiments with
longer pulses \#1 and \#2 also had a higher fraction of missing
atoms. For example, when we used 50~$\mu$s pulses with $n_0 = 5.4
\times 10^{13}$~cm$^{-3}$, 56(3)\% were missing.

We have carried out double-pulse measurements with a variety of
widths and amplitudes for pulses \#1 and \#2 and a variety of
different densities and initial magnetic fields. Although the
oscillation frequency was unchanged, the phase, contrast, and
damping of the oscillations did vary. The contrast was very
sensitive to the pulse length, and was lower for longer pulses
that created more missing atoms. Defining the contrast as the
oscillation amplitude divided by the time averaged number of
remnant atoms detected, we observed an optimum contrast of 0.42(7)
for 15~$\mu$s pulses to 156.6(1)~G. A single such pulse removed
about half of the atoms from the BEC. When the pulse length was
comparable to $\tau_{decay}$, the contrast was reduced by about a
factor of two. Under those conditions, $\sim$3/4 of the atoms were
lost after pulse \#1. The phase was shifted, but we did not
observe a change in the contrast when we varied the amplitudes of
pulses \#1 and \#2 from $B$ = 156.6(1)~G (2,400~$a_0$) to $B$ =
155.1(1)~G (24,000~$a_0$). The contrast did depend on the
intermediate level, however, and was reduced for $B_{evolve}$
values closest to the resonance, for which the magnetic-field
jumps between $B_{evolve}$ and pulses \#1 and \#2 were shortest.

We also looked for a temperature dependence of both the damping
and the frequency at $\sim$200~KHz and did not see any. The
high-temperature data was much noisier than the data for pure
condensates,  due to unexplained enhanced noise in the number of
thermal atoms after the magnetic-field pulse, but when the initial
thermal fraction was increased from $<$5\% to 30\%, the data still
fit to oscillations with frequency, amplitude, and damping
consistent with what was observed with low temperature data.

{\bf Conclusions and Outlook}. Our interpretation of our
observations is that the first magnetic-field pulse provides a
sufficiently rapid perturbation to result in nonadiabatic mixing
between atomic and molecular states. The superposition then
evolves according to the energy difference between the states,
which is determined by the magnetic field during the free
evolution stage, $B_{evolve}$. The second pulse mixes atom and
molecule states again, such that the final state of the system
depends on the relative phase of atomic and molecular fields at
the time of the second pulse. This is somewhat analogous to
Ramsey's method of separated oscillating fields$^{33}$. Under very
limited conditions ($a$ near 1700~$a_0$), we could also observe
Rabi-like oscillations with a single pulse towards the Feshbach
resonance. This narrow window results from the conflicting needs
for both strong coupling and condensate loss time$^{25}$ long
compared to a Rabi oscillation period.

After pulse \#2, a fraction of the coherent molecular component is
converted into the energetic but still spin-polarized burst atoms
through a yet to be determined process. Another mystery is the
missing atoms. Are they molecules that are not converted back into
atoms and are not detected in the burst or the remnant signals? If
so, why do we not see them as atoms after the field is turned off
and the corresponding molecular state is no longer bound?  Why are
there fewer missing atoms for lower-density condensates and
quicker pulses towards the Feshbach resonance? What is the actual
conversion efficiency from atoms to molecules and how could we
maximize it? Very near the Feshbach resonance, the molecular state
has a magnetic moment nearly the same as that of the free atoms,
and hence will remain magnetically trapped. A major remaining
question concerns the nature of the molecules. Could they be
considered a molecular BEC? Clearly there is much to be learned
about this curious system.

\begin{itemize}
\item[1.] Wynar, R.~H., Freeland, R.~S., Han, D.~J., Ryu, C. \&
Heinzen, D.~J. Molecules in a Bose-Einstein condensate. {\em
Science} {\bf 287}, 1016-1019 (2000).
\item[2.] Heinzen, D.~J., Wynar, R., Drummond, P.~D. \&
Kheruntsyan, K.~V. Superchemistry: dynamics of coupled atomic and
molecular Bose-Einstein condensates. {\em Phys. Rev. Lett.} {\bf
84}, 5029-5033 (2000).

\item[3.] Anglin, J.~R. \&  Vardi, A. Dynamics of a two-mode Bose-Einstein
condensate beyond mean-field theory. {\em Phys. Rev. A} {\bf 64},
013605/1-9 (2001).

\item[4.] Cusack, B.~J., Alexander, T.~J., Ostrovskaya, E.~A. \&
Kivshar, Y.~S. Existence and stability of coupled atomic-molecular
Bose-Einstein condensates. {\em Phys. Rev. A} {\bf 65}, 013609/1-4
(2001).

\item[5.] Calsamiglia, J., Mackie, M. \& Suominen, K. Superposition of
macroscopic numbers of atoms and molecules. {\em Phys. Rev. Lett.}
{\bf 87}, 160403-1 (2001).

\item[6.] Drummond, P.~D., Kheruntsyan, K.~V., Heinzen, D.~J. \& Wynar, R.~H. Stimulated Raman adiabatic passage from an atomic to a molecular
Bose-Einstein condensate. (available at
http://lanl.arxiv.org/abs/cond-mat/0110578) 1-16 (2002).
\item[7.] McKenzie, C. {\em et~al.} Photoassociation of Sodium in a Bose-Einstein Condensate.
{\em Phys. Rev. Lett.} {\bf 88}, 120403/1-4 (2001).

\item[8.]
Tiesinga, E., Moerdijk, A., Verhaar, B.~J. \& Stoof, H.~T.~C.
Conditions for Bose-Einstein condensation in magnetically trapped
atomic cesium. {\em Phys. Rev. A} {\bf46}, R1167-R1170 (1992).

\item[9.]
Tiesinga, E., Verhaar, B.~J. \& Stoof, H.~T.~C. Threshold and
resonance phenomena in ultracold ground-state collisions. {\em
Phys. Rev. A} {\bf 47}, 4114-4122 (1993).

\item[10.]
Moerdijk, A.~J., Verhaar, B.~J. \& Axelson, A. Resonances in
ultracold collisions of $^{6}$Li, $^{7}$Li, and $^{23}$Na. {\em
Phys. Rev. A} {\bf 51}, 4852-4861 (1995).

\item[11.]
van Abeelen, F.~A. \& Verhaar, B.~J. Time-dependent Feshbach
resonance scattering and anomalous decay of a Na Bose-Einstein
condensate.
 {\em Phys. Rev. Lett.} {\bf 83},
1550-1553 (1999).

\item[12.]
Mies, F.~H., Tiesinga, E. \& Julienne, P.~S. Manipulation of
Feshbach resonances in ultracold atomic collisions using
time-dependent magnetic fields. {\em Phys. Rev. A} {\bf 61},
022721/1-17 (2000).

\item[13.]
van Abeelen, F.~A., Heinzen, D.~J. \& Verhaar, B.~J.
Photoassociation as a probe of Feshbach resonances in cold-atom
scattering. {\em Phys. Rev. A} {\bf 57}, R4102-R4105 (1998).

\item[14.] The coupled--channels calculations were
kindly provided by Servaas Kokkelmans and Chris Greene. The
calculations used the best estimates of the Feshbach resonance
parameters found by combining the results of three high-precision
experiments. The determination of the parameters and the
construction of the potentials is described in detail by van
Kempen et al.$^{15}$

\item[15.] van Kempen, E.~G.~M., Kokkelmans, S.~J.~J.~M.~F.,
Heinzen, D.~J. \& Verhaar, B.~J. Interisotope determination of
ultracold rubidium interactions from three high-precision
experiments. {\em Phys. Rev. Lett.} {\bf 88}, 093201/1-4 (2002).

\item[16.]  Timmermans, E., Tommasini, P., Hussein, M. \&
Kerman, A. Feshbach resonances in atomic Bose-Einstein
condensates.
 {\em Phys. Rep.} {\bf 315}, 199-230 (1999).

\item[17.] Timmermans, E., Tommasini, P., C\^{o}t\'{e}, R., Hussein, M. \&
Kerman, A. Rarified liquid properties of hybrid atomic-molecular
Bose-Einstein condensates. {\em Phys. Rev. Lett.} {\bf 83},
2691-2691 (1999).

\item[18.] Drummond, P.~D., Kheruntsyan, K.~V. \& He, H.
Coherent molecular solitons in Bose-Einstein condensates. {\em
Phys. Rev. Lett.} {\bf 81}, 3055-3058 (1998).

\item[19.] Holland, M., Park, J. \& Walser, R. Formation of pairing fields in
resonantly coupled atomic and molecular Bose-Einstein condensates.
{\em Phys. Rev. Lett.} {\bf 86}, 1915-1918 (2001).

\item[20.] G\'{o}ral, K., Gajda, M. \& Rz\c{a}\.{z}ewski, K.
Multimode dynamics of a coupled ultracold atomic-molecular system.
{\em Phys. Rev. Lett.} {\bf 86}, 1397-1401 (2001).

\item[21.] Vardi, A., Yurovsky, V.~A. \& Anglin, J.~R.
Quantum effects on the dynamics of a two-mode atom-molecule
Bose-Einstein condensate. {\em Phys. Rev. A} {\bf 64}, 063611/1-5
(2001).

\item[22.] Stenger, J. {\em et~al.} Strongly enhanced inelastic collisions in a Bose-Einstein
condensate near Feshbach resonances. {\em Phys. Rev. Lett.} {\bf
82}, 2422-2425 (1999).

\item[23.] Cornish, S.~L., Claussen, N.~R., Roberts, J.~L., Cornell, E.~A. \&
Wieman, C.~E. Stable $^{85}$Rb Bose-Einstein condensates with
widely tunable interactions. {\em Phys. Rev. Lett.} {\bf 85},
1795-1798 (2000).

\item[24.] Claussen, N.~R., Cornish, S.~L., Roberts, J.~L., Cornell, E.~A. \&
Wieman, C.~E. in {\em Atomic Physics 17}, (ed. Arimondo, E.,
DeNatale, P., \& Inguscio, M.) 325-336 (American Institute of
Physics, New York, 2001).

\item[25.]  Claussen, N.~R., Donley, E.~A.,
Thompson, S.~T. \& Wieman, C.~E. Microscopic Dynamics in a
Strongly Interacting Bose-Einstein Condensate. in press (available
at http://lanl.arxiv.org/abs/cond-mat/0201400) 1-4 (2002).

\item[26.] Roberts, J.~L. {\em et~al.} Controlled collapse of a Bose-Einstein condensate.
{\em Phys. Rev. Lett.} {\bf 86}, 4211-4214 (2001).

\item[27.] P{\'e}rez-Garc{\'i}a, V.~M., Michinel, H., Cirac, J.~I., Lewenstein, M. \&
Zoller, P. Dynamics of Bose-Einstein condensates: variational
solutions of the Gross-Pitaevskii equations. {\em Phys. Rev. A}
{\bf 56}, 1424-1432 (1997).

\item[28.] Donley, E.~A.  {\em et~al.} Dynamics of collapsing and exploding Bose-Einstein condensates.
{\em Nature} {\bf 412}, 295-299 (2001).

\item[29.] Sakurai, J.~J. {\em Modern Quantum Mechanics}
(Addison-Wesley, Reading, Massachusetts, 1994).

\item[30.] Rethermalization measurements in $^{85}$Rb thermal
clouds yield values of $B_0$ = 154.9(4)~G and $\Delta$ = 11.0(4)~G
(31). The best estimate of $a_{bg}$ is an estimate found by mass
scaling spectroscopic measurements of the four highest-lying bound
states of $^{87}$Rb$^{15}$, which gives $a_{bg} =
-450(3)~a_0$$^{32}$.

\item[31.] Roberts, J.~L. {\em et~al.}  Improved characterization of elastic scattering near a Feshbach resonance
in $^{85}$Rb. {\em Phys. Rev. A} {\bf 64}, 024702/1-3 (2001).

\item[32.] Kokkelmans, S.~J.~J.~M.~F. Private Communication.

\item[33.] Ramsey, N.~F. A molecular beam resonance method with
separated oscillating fields. {\em Phys. Rev.}
{\bf 78}, 695-699 (1950).
\end{itemize}

\noindent{\bf Acknowledgements}

\noindent We would like to acknowledge contributions from E.~A.
Cornell and the JILA quantum gas collaboration. We are
particularly grateful to C.~H. Greene and S.~J.~J.~M.~F.
Kokkelmans for providing us with the coupled-channel scattering
calculations presented in Fig. 4 and to L.~Pitaevskii for numerous
fruitful discussions. S.~T.~T. acknowledges the support of an
ARO-MURI Fellowship.  This work was also supported by ONR and NSF.

\vspace{1 cm}

\noindent Correspondence and requests for materials should be
addressed to N.R.C.

\noindent (e-mail:nclausse@jilau1.colorado.edu.)

\end{document}